\documentclass[12pt]{iopart}

\usepackage{graphicx}
\begin{document}

\title{Axial-vector weak coupling at medium momentum for astro neutrinos and double beta decays }

\author[first]{H. Ejiri}
\address{Research Center for Nuclear Physics, 
Osaka University, Osaka 567-0047, Japan} 
\ead{ejiri@rcnp.osaka-u.ac.jp} 
\vspace{10pt}
\begin{abstract}
Neutrino nuclear responses associated with medium momentum transfer of $q$=20-80 MeV/c 
for astro neutrinos and double beta decays (DBDs) were studied by using charge exchange  ($^{3}$He,$t$) reactions on 
$^{128}$Te and $^{130}$Te.   Gamow-Teller  and spin-dipole nuclear matrix elements (NMEs) are found to be reduced uniformly 
in the wide momentum region of $q$=20 - 80 MeV/c 
with respect to pnQRPA NMEs by the coefficient of $k_{NM}\approx$ 0.35, while the Fermi NME for the isobaric analogue state is 
given by the sum rule limit in the momentum transfer region.
The reduction is discussed in terms of the quenching of the axial vector coupling $g_A^{eff}$ for astro-neutrinos and DBDs.
.\\

nuclear matrix element, momentum transfer, quenching of axial vector transitions, charge exchange reaction.

\end{abstract}

\section{Introduction}

Neutrino nuclear responses for astro neutrinos and double beta decays (DBDs) 
 are crucial for studying neutrino properties of astro and particle physics interests. 
They are discussed in review articles and references therein \cite{eji78,eji00}. Recent works on DBD experiments 
and the nuclear matrix elements (NMEs)
are given in 
reviews and references therein  in 

The neutrino response for the axial vector weak coupling is of current interest in views of possible renormalization (quenching) effect. The responses
 studied by $\beta-\gamma $ transitions at the low momentum transfer of $q \approx$a few MeV/c
are reduced with respect to simple model evaluations. On the other hand 
the responses involved in  supernova neutrinos and neutrino-less DBDs are associated with the medium momentum transfer of
  $q\approx$ 20-100 MeV/c.
  
The present work aims to study the momentum dependence of  the axial-vector neutrino responses relevant to such astro neutrinos
and DBDs  by using charge exchange reactions (CERs), and to
show that the renormalization (quenching) factors for the Gamow-Teller (GT) 
and spin-dipole (SD) responses are universal in the wide $q$-transfer region of $q$=2-100 MeV. 

The neutrino (weak) nuclear response is 
expressed in terms of the weak coupling $g_{W}$  and the NME $M(\alpha)$ for the $\alpha$ mode interaction as 
\begin{equation}
g_{\rm W} ^2B(\alpha) = g_W^2(2J_i+1)^{-1} |M(\alpha)|^2, 
\end{equation}
where $J_i$ is the initial state spin, and $g_{W}$ is the nucleon weak coupling with W being V for the vector coupling and A for 
the axial vector one.  
The interaction modes to be discussed in the present report are 
$\alpha$=F (Fermi), GT  and SD.  The F and GT modes are involved mainly
 in low-energy astro neutrinos and the GT mode in the two neutrino DBDs. On the other hand 
the SD mode is also involved in the higher-energy supernova neutrinos and is one of the major components of the neutrino-less DBDs \cite{suh12}. 
 
The weak couplings for the momentum transfer $q$ are usually expressed by using the
dipole approximation,
\begin{equation} 
\label{eq:dipole}
   g_{\rm W} = \frac{g_{\rm W}^0}{\big(1+q^2/M_{\rm W}^2\big)^2},
 \end{equation}
where $g_{\rm W}^0$ is the weak
coupling strength at zero momentum transfer ($q^2=0$), and $M_{\rm W}$ is the weak mass, respectively. For the vector and 
axial-vector masses one usually takes $M_{\rm V}=840\,\textrm{MeV}$ 
and $M_{\rm A}\sim 1\,\textrm{GeV}$ \cite{bod08,bha11,ama15} coming
from the accelerator-neutrino phenomenology. In the present momentum region of $q\approx $20-100 MeV/c, the $q$ dependence 
is very small, only of an order of 0.02 or less. Then one may assume constant $g_{W}$ in the present momentum region. Then the $q$ dependence 
of the response is considered to be mainly concerned with that for the NMEs $M(\alpha)$.

 So far, the GT and SD responses (NMEs) have been studied experimentally by investigation
 $\beta $ decays, electron captures (ECs) and $\gamma $ decays, 
where the momentum transfer involved is as low as $q\le$ a few MeV/c.

 \section{Charge exchange reactions on $^{128}$Te and $^{130}$Te}
 
Recently high energy-resolution CER experiments  of $(^3$He, $t$) have been studied for nuclei of astro-neutrino and DBD interests at RCNP Osaka University. 
as given in the review articles \cite{eji00,eji05,ver12}. 
The Fermi, GT and SD responses are derived by analyzing the CERs in the angular range of $\theta$=0-4 deg., corresponding to the 
momentum range  of 5-120 MeV/c. 

 In the present work, we discuss F, GT and SD responses  for even even nuclei with $J_i$=0. Then the spin factor is  2$J_i$+1=1. 
The interaction operators are given as 
\begin{equation}
T(F)=\tau^{\pm},~~ T(GT)= \tau^{\pm} \sigma ~~  T(SD)=\tau ^{\pm} [i^1\sigma \times f(r) Y_1]_2 , 
\end{equation}
where $\tau $ and $\sigma$ are the isospin and spin operators, respectively. 
The NMEs are given as $M(\alpha)=<T(\alpha)>$ with $\alpha$=F, GT, and SD.
Actually, there are strong isospin  and spin  nuclear interactions, 
which are repulsive in nature, and thus the  NMEs are much modified in nucleus due to 
nucleonic and non-nucleonic isospin spin correlations \cite{eji00,ver12}.

Then the  NME  is given as 
\begin{equation}
M(\alpha) =  k^{eff}_{NM}M_{QR}(\alpha), 
\end{equation}
where $M_{QR}(\alpha)$ is
 the pnQRPA model NME and $k^{eff}_{NM}$ stands for the re-normalization (quenching) coefficient due to 
  all kinds of nuclear and non-nuclear correlation and nuclear medium effects,  
 which are not explicitly included in the pnQRPA model NME.

Recent analyses of $\beta^- $ and EC NMEs at the low-momentum transfer of $q \le$ a few MeV/c show
 i. the F strength is concentrated in one isobaric analog state (IAS) and  no strengths are located in the low-lying states, and ii.  
the GT and SD strengths at the low-excitation region are shifted up to the broad GT and SD giant resonances (GRs) at the high excitation region, and thus
 the GT and SD NMEs for low-lying states  are reduced much from the 
simple quasi-particle pnQRPA NMEs $M_{QR}(\alpha)$ by the reduction factor of the order of $k_{NM}(\alpha)\approx$0.4-0.6 
partly due to the non-nucleonic $\tau\sigma $ correlations and nuclear medium effects \cite{eji78,eji00,eji15,eji14}.
 
In the present report we discuss the $q$ dependence of  CERs on $^{128,130}$Te , which are of astro-neutrino and DBD interest, for 
the F (IAS), the lowest GT state and the lowest SD state in $^{128,130}$I \cite{pup12,eji16}. 
The GT and SD states are mainly excitation  of the 
single quasi-neutron to single quasi-proton states of the 2d(3/2)$_n\rightarrow$ 2d(5/2)$_p$ and  
the 1h(11/2)$_n \rightarrow$(1g7/2)$_p$, while the F (IAS) state is the Fermi GR for the coherent $j=(l\pm1/2)_n\rightarrow j=(l\pm1/2)_p$ excitations. 

The $Q$ values of  the CERs for the
 low-lying states are much smaller than the incident $^3$He beam energy. Thus the momentum transfer 
is given by $q\approx P_i \theta$ with $P_i$ and $\theta$ being the incident $^3$He momentum and the out-going $t$ angle. 

The reaction diagram is schematically shown in Fig. 1. 

\begin{figure}[h]
\begin{center} 
\includegraphics[width=0.4\textwidth]{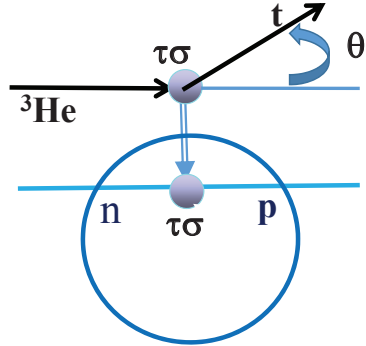}
\end{center} 
\caption{ Schematic diagram of the ($^3$He,$t$) for the isospin ($\tau$) spin ($\sigma$) excitation. $\theta$ denotes the scattered angle of $t$.
\label{fig:fig1}}
\end{figure}

In case of the RCNP beam energy of $E$=0.42 GeV, 
the differential cross sections in the medium momentum region of $q\approx$ 20-100 MeV/c 
are measured by observing  the angular distribution of $t$ over $\theta$=0-4 deg. 

The differential cross sections for them are derived as a function of the 
momentum transfer $q$ from the observed angular distributions \cite{pup12}.  The obtained $q$ distributions for the F, GT and SD states in  $^{128}$I and 
 $^{130}$I  are shown, respectively,  in Fig. 2 and Fig. 3.
 
The differential cross section for the $\alpha$ state is expressed as
\begin{equation}
\frac{d\sigma(\alpha)}{d\Omega}=K(\alpha)F(\alpha,q) J(\alpha)^2 B(\alpha), 
\label{equ:equ2321}
\end{equation}
\begin{equation}
B(\alpha)= \kappa ^{eff}(q)^2B^0(\alpha),
\label{equ:equ2322}
\end{equation}
where $K(\alpha)$ and $J(\alpha)$ with $\alpha$=F, GT, and SD are the kinematic factors and the 
volume integrals of the interaction, respectively.  The kinematic $q$-dependence is given by  
 $F_i(\alpha,q)$. The $q$-dependent response is effectively expressed as  $B(\alpha)=
\kappa ^{eff}(q)^2$ $B^0(\alpha)$ with $B^0(\alpha)$ being the nuclear response at $q$=0. The coefficient $k^{eff}(q)$ stands for the 
effective $q$-dependent coupling.  Accordingly, the NMEs are given as $M(\alpha)=
\kappa ^{eff}(q)$ $M_i^0(\alpha)$ with $M_i^0(\alpha)$ being the NME at $q$=0.

\begin{figure}[t]
\begin{center} 
\includegraphics[width=1.0\textwidth]{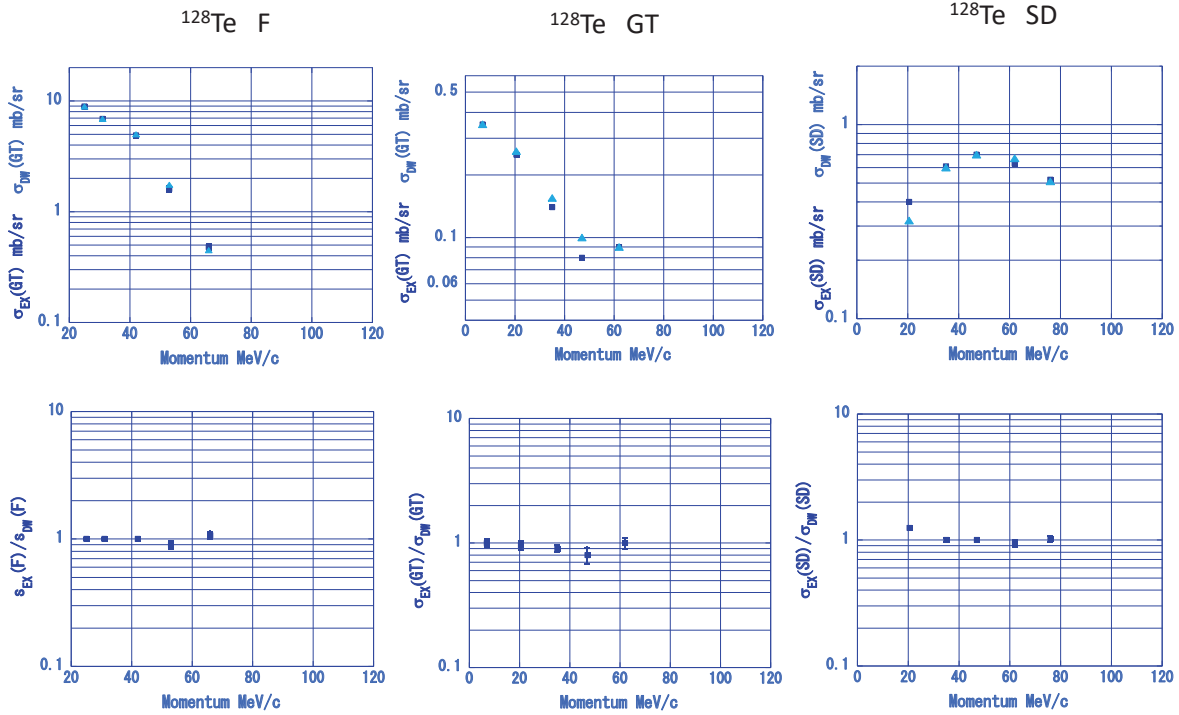}
\end{center} 
\caption{ Top: The $^{128}$Te ($^3$He,$t$)  CER cross sections as a function of the momentum transfer $q$.  
 F, GT and SD stand for the  0$^+$ 8.31 MeV IAS, the 1$^+$ 0.12 MeV GT state,  and the  2$^-$ ground SD state, respectively.  
 Squares: Experimental $\sigma_{EX}(\alpha)$. Triangles: DWBA $\sigma_{DW}(\alpha)$.
Bottom:  Squares: The ratio  of the experimental to the  DWBA cross sections as a function of the momentum transfer $q$
 for $\alpha$= F (IAS), GT (1st GT), and SD (ground) states.   See text. 
\label{fig:fig2}}
\end{figure}

\begin{figure}[t]
\begin{center} 
\includegraphics[width=1.0\textwidth]{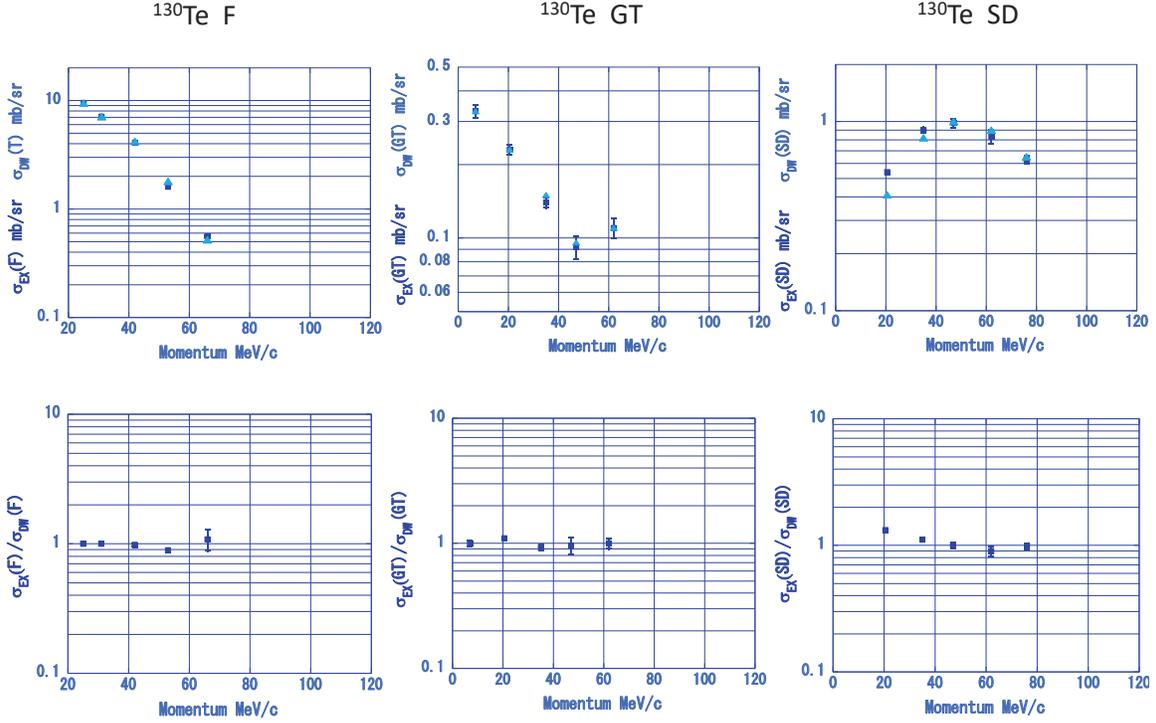}
\end{center} 
\caption{ Top: The $^{130}$Te ($^3$He,$t$) CER cross sections as a function of the momentum transfer $q$.  
 F, GT, and SD stand for the  0$^+$: 12.72 MeV IAS,  the 1$^+$ 0.12 MeV GT state and the 2$^-$  ground SD state, respectively.  
The solid lines are the DWBA calculations. Squares: Experimental values. Triangles: DWBA values.
Bottom:  Squares: The ratio  of the experimental to the  DWBA cross 
sections as a function of the momentum transfer $q$
 for $\alpha$= F (IAS), GT (1st GT), and SD (ground) states.  See text. 
\label{fig:fig3}}
\end{figure}

The kinematic $q$-dependence $F_i(\alpha,q)$ is evaluated by the DWBA calculation.
 In the present case of the medium energy projectile, the distortion effect is rather small, and thus the kinematic factor is approximately given by 
 the square of the spherical Bessel function of $|J_l(qR)|^2$ with $l$  being the orbital angular momentum transfer, and $R$ being the effective interaction radius. 
 The $q$-dependent coupling
 $\kappa ^{eff}(q)$ manifests as deviation of the observed $q$ (angular) distribution from the DWBA calculation. The 
 observed $q$-dependences of the CER cross sections $\sigma _{EX}(\alpha)/d\Omega$ for $\alpha$=F, GT, and SD responses
 are well reproduced by the DWBA calculations $\sigma _{DW}(\alpha)/d\Omega$ with constant $\kappa ^{eff}(q)^2$ as 
shown in Figs. \ref{fig:fig2} and \ref{fig:fig3}. 

The ratios of 
$\sigma _{EX}(\alpha)/d\Omega $  to the $\sigma _{DW}(\alpha)/d\Omega $ for $^{128}$Te and $^{130}$Te are also shown in Figs. 2 and 3. 
We note that the ratios are nearly 1 for the momentum region of the present interest. Accordingly the experimental distributions for F, GT, and SD 
components are well expressed by the DWBA ones with the kinematic $q$-dependent distribution and the  constant coefficient for 
each  $\kappa ^{eff}(\alpha)$ for the $\alpha$= F, GT, and SD NMES over the momentum region of $q$=30-80 MeV/c .   

Actually, the observed and DWBA $q$-distribution for the 0$^+$ F (IAS) states consists of the pure $l$=0  component, 
while  those  for the 1$^+$  GT ($l$=0)  include the admixture of the $l=2$ component beyond the momentum $q\ge$ 40 MeV, and 
and those for the 2$^-$ SD ($l$=1) state include the admixtures of the $l$=3  component beyond the momentum $q\ge$60 MeV.  
In other words, the good agreement of the experimental and  DWBD $q$ distributions in the wide momentum region of $q$=10-80 MeV/c, which is 
relevant to the angular momentum transfers of $\l$=0-3 suggest the universal quenching coefficients over the wide linear and angular momentum regions
of the astro and DBD interests.  The present results are in accord with the discussion on $^{76}$Ge in the review \cite{eji19}. 

\section{Quenching of the axial-vector weak coupling}
 
 The GT and SD NMEs at the low $q$ ($\le$ a few MeV) region for DBD and other medium-heavy nuclei are experimentally available 
 from the $\beta$/EC data. 
They are reduced with respect to the pnQRPA NMEs by the coefficient $k_{NM}(0)\approx$0.35-0.65 at the 
$\beta $/EC point of $q \approx$0 \cite{eji15,eji14}. The experimental GT and SD NMEs in the Te isotopes are shown in the left panel of Figure 4, and the 
re-normalization (quenching) coefficients for the GT and SD NMEs with respect to the QRPA NMEs are shown in the right panel. 
They are quite uniform in the mass region of A=122-130. 
Thus the axial-vector weak coupling is considered to be uniformly re-normalized (quenched) by the 
coefficient $k_{NM}(q) \approx $0.35 in the wide momentum region of $q$=0-80 MeV/c, which is 
the region of the neutrino-less DBDs and medium-energy supernova neutrinos . 

\begin{figure}[t]
\begin{center} 
\includegraphics[width=0.7\textwidth]{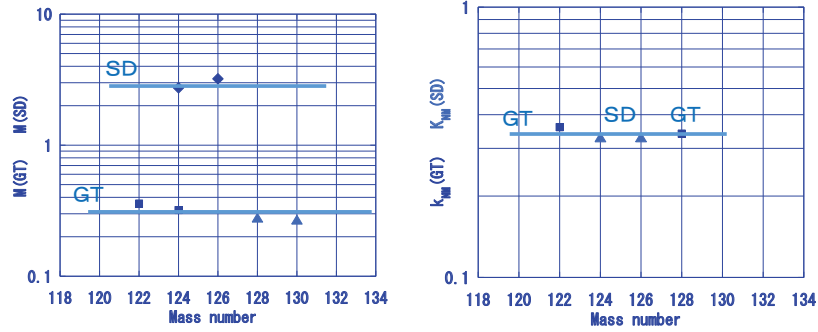}
\end{center} 
\caption{Left side panel: NMEs for GT and SD transitions in Te isotopes.  Squares triangles and diamonds are the GT NMEs from
$\beta$/EC decays, the GT NMEs from CERs and the SD NMES in units of 10$^{-3}$ n.u. from the $\beta $/EC decays, respectively.  Right side panel: Re-normalization
(quenching) coefficients  for GT and SD NMEs. Squares and triangles are the values for the GT $\beta$/EC  and the SD $\beta$/EC decays, respectively.   See text.  
\label{fig:fig4}}
\end{figure}

\begin{table}[htb] 
\caption{Experimental and pnQRPA GT and SD NMEs at $q\approx$ a few MeV for Te isotopes. 
SD NMEs are in unit of 10$^{-3}$ n.u.  See refs.[14,15].   $k^{eff}_{NM} $: effective reduction factor. 
See tex}
\vspace{0.5 cm}
\centering
\begin{tabular}{ccccc}
\hline
Transition  & $\alpha$ & $M_{EX}(\alpha)$ &  $M_{QR}(\alpha)$ & $k^{eff}_{NM}$\\
\hline
$^{124}$Te $\leftrightarrow^{124}$I& GT   & 0.36 &  0.99 & 0.36\\
$^{126}$Te$\leftrightarrow^{126}$I & SD    & 2.743 &  8.38 & 0.33\\
$^{128}$Te$\leftrightarrow^{128}$I & SD    & 3.22&  9.62 & 0.33\\
$^{130}$Te $\leftrightarrow^{130}$I & GT   &0.32 &  0.90 & 0.34\\

\hline
\end{tabular}
\end{table} 

The present work shows the axial-vector GT and SD NMEs are reduced (quenched) with respect to the pnQRPA NMEs in the wide momentum region of the neutrino-less DBD and super-nova neutrino interest as seen in the GT and SD NMEs at the low momentum region found in the $\beta $/EC decays 
\cite{eji15,eji14}.  The NMEs studied in the present CERs are the $\tau^-$-side NMEs. 

The $\tau^+$-side neutrino response has recently been studied by ordinary muon capture (OMC) reaction \cite{has18,zin19,jok19}. The 
axial-vector NMEs in the similar 
momentum region of $q$=50-100 MeV/c are found to be reduced by the re-normalization coefficient of $k_{NM}\approx$0.4 with  respect to the pnQRPA NMEs.
\cite{has18,jok19}.  Then, one may expect the DBD NMEs are reduced by a re-normalization coefficient around $k_{NM}\approx$0.3-0.4 with respect to the pnQRPA NMEs, depending on the 
relative weight of the axial-vector NME. 

The supernova neutrino is in the medium momentum region of 10-50 MeV/c. Then the NME for the axial vector transition is given by
\begin{equation}
 M^{\nu}=k_{NM} M^{\nu}_{QR}(\alpha),
 \end{equation}
  where  $ k_{NM}\approx$ 0.3-0.4 is the re-normalization (quenching) coefficient and  $ M^{\nu}_{QR}(\alpha)$ is the pnQRPA NME. Then the interaction rate is reduced by the coefficient of $(k_{NM})^2\approx$ 0.1-0.2 with respect to the pnQRPA rate for the wide $q$ region.

The neutrino-less DBD NME is given by 
\begin{equation}
M^{0\nu}=(k^{eff}_{NM})^2[M^{0\nu}_{GT} + M^{0\nu}_{T}] + M^{0\nu}_{F}],
\end{equation}
where $M^{0\nu}_{GT},  M^{0\nu}_{T}$ and  $M^{0\nu}_{F}$ are the axial-vector, tensor, and vector DBD NMEs, respectively. Then the axial-vector and tensor NMEs 
are reduced with respect to the pnQRPA NMEs by the re-normalization (quenching) coefficient of $(k^{eff}_{NM})^2\approx$ 0.1-0.2, and the DBD NME by the re-normalization (quenching) coefficient of 0.15-0.3, depending on the relative weight of the vector NME with respect to the axial-vector + tensor NMEs.  
Then the DBD isotope (detector) mass required for a given $\nu$-mass sensitivity is 2-3 orders of magnitude more that the detector mass in case of the pnQRPA
MNE.

In case of the two neutrino $\beta \beta $  NMEs , the matrix element is given by the product of the $\tau^-$-side and $\tau^+$-side GT NMEs at
the low  $q$=0-3 MeV/c Thus the two-neutrino NMEs are found to be reduced by the coefficient $k\approx k^- \times k^+$ with 
$k^{\pm}$ being the reduction coefficient for the $\tau^{\pm}$ GT NMEs \cite{eji19,eji09,eji12}.  Accordingly, the axial-vector DBD NME might be reduced by the similar coefficient.

It is noted that the re-normalization (quenching) coefficient $k_{NM}$ is conventionally expressed as $g_A^{eff}/g_A$.  It is very interesting to evaluate theoretically the re-normalization coefficient by taking into accounts all kinds of nuclear medium and non-nucleonic correlation effects, which are not explicitly included in the pnQRPA model. \\

The author thank Prof. J. Suhonen for valuable discussions.\\

{\bf References}\\

\end{document}